\documentclass[11pt,a4paper]{article}
\usepackage{ae,aecompl}
\usepackage[T1]{fontenc}
\usepackage[latin9]{inputenc}
\usepackage{color}
\usepackage{textcomp}
\usepackage{amsmath}
\usepackage{amssymb}
\usepackage{graphicx}
\usepackage{esint}

\makeatletter

\newcommand*\LyXZeroWidthSpace{\hspace{0pt}}

\pdfoutput=1 

\usepackage{jheppub}

\usepackage{etoolbox}
    
    \patchcmd{\maketitle}{\@fpheader}{}{}{}

\usepackage{aecompl}

\usepackage{amsfonts}

\usepackage{bbm}

\title{\boldmath BTZ black hole with KdV-type boundary conditions: Thermodynamics revisited}

\author[a,b]{Cristi\'{a}n Erices,}
\author[c,d,e]{Miguel Riquelme,}
\author[c,f,g]{Pablo Rodr\'{i}guez,}

\affiliation[a]{Department of Physics, National Technical University of Athens, Zografou Campus GR 157 73, Athens, Greece.}
\affiliation[b]{Universidad Cat\'{o}lica del Maule, Av. San Miguel 3605, Talca, Chile.}
\affiliation[c]{Centro de Estudios Cient\'{i}ficos (CECs), Av. Arturo Prat 514, Valdivia, Chile.}
\affiliation[d]{Facultad de Ingenier\'{i}a y Tecnolog\'{i}a, Universidad San Sebasti\'{a}n, General Lagos 1163, Valdivia 5110693, Chile.}
\affiliation[e]{Fundación Cultura Cient\'{i}fica, Valdivia 5112119, Chile.}
\affiliation[f]{Departamento de F\'{i}sica, Universidad de Concepci\'{o}n, Casilla 160-C, Concepci\'{o}n, Chile.}
\affiliation[g]{Instituto de Ciencias F\'{i}sicas y Matem\'{a}ticas, Universidad Austral de Chile, Casilla 567, Valdivia, Chile}

\emailAdd{crerices@central.ntua.gr}
\emailAdd{riquelme@cecs.cl}
\emailAdd{rodriguez@cecs.cl}

\abstract{The thermodynamic properties of the Bañados-Teitelboim-Zanelli (BTZ) black hole endowed with Korteweg-de Vries (KdV)-type boundary conditions are considered. This familiy of boundary conditions for General Relativity on AdS$_{3}$ is labeled by a non-negative integer $n$, and gives rise to a dual theory which possesses anisotropic Lifshitz scaling invariance with dynamical exponent $z=2n+1$. We show that from the scale invariance of the action for stationary and circularly symmetric spacetimes, an anisotropic version of the Smarr relation arises, and we prove that it is totally consistent with the previously reported anisotropic Cardy formula. The set of KdV-type boundary conditions defines an unconventional thermodynamic ensemble, which leads to a generalized description of the thermal stability of the system. Finally, we show that at the self-dual temperature $T_{s}= \frac{1}{2\pi}(\frac{1}{z})^{\frac{z}{z+1}}$, there is a Hawking-Page phase transition between the BTZ black hole and thermal AdS$_{3}$ spacetime.}

\makeatother

\begin{document}
\maketitle \flushbottom

\section{Introduction}

In the pursuit of a better understanding of quantum gravity, in the
past two decades, a lot of interest has been put into the so called
Gauge/Gravity correspondence, whose most celebrated example is the
AdS/CFT duality \cite{Maldacena,Witten-1}. In this context, AdS$_{3}$/CFT$_{2}$
correspondence has played an important role. One of the first main
results was the renowned article from Brown and Henneaux \cite{BrownHenneaux},
where they showed that the asymptotic symmetries of General Relativity
in three dimensions with negative cosmological constant correspond
to the conformal algebra in two dimensions with a classical central
extension. This result naturally suggest that a quantum theory of
gravity in three dimensions could be described by a CFT at the boundary.
Based on this result, Strominger \cite{Strominger:1997eq} proved
that the entropy of the Bañados-Teitelboim-Zanelli (BTZ) black hole
\cite{BTZ,BHTZ} can be recovered by a microscopic counting of states
by means of the Cardy formula \cite{Cardy}. This simple example gave
rise to an active field of research regarding the thermodynamic properties
of lower dimensional black holes and how they could be holographically
related to a dual field theory that describes the much sought after
quantum theory of gravity.

In the static case, the thermodynamic stability of the BTZ black hole
has qualitatively a different behaviour than its higher dimensional
counterparts. In fact, in four dimensions, the canonical ensemble
for the \textcolor{black}{Schwarzschild solution} is not well-defined
\cite{Hawking}\textcolor{black}{. This difficulty is avoided}\textcolor{red}{{}
}\textcolor{black}{by the presence of the negative cosmological constant
\cite{HawkingPage}, }but nevertheless\textcolor{black}{, since the
specific heat of Schwarzschild-AdS$_{4}$ black hole presents discontinuities,
the system can not reach thermal equilibrium with a thermal bath at
any temperature. In the case of three dimensions, none of the above
arguments hold \cite{Zaslavskii,Mann-Brown,Reznik}, since the specific
heat of the BTZ is a monotonically increasing positive function of
the temperature, as the case of any Chern-Simons black hole in odd
dimensions \cite{Muniain,Crisostomo,Cai}.}

Several efforts have been made to generalize the Gauge/Gravity proposal
for non AdS asymptotics (see e.g. \cite{Strominger,Guica,Barnich,Detournay}).
In this scenario, a lot of attention has been placed on gravity dual
theories with anisotropic scaling properties, which are found in the
context of non-relativistic condense matter physics (see references
in \cite{CMT duals}). The main work on this subject has been done
along the lines of \emph{Lifshitz holography}, where the gravity counterparts
are given by asymptotically Lifshitz geometries (see e.g. \cite{Marika}
and references therein). However, this class of spacetimes are not
free of controversies. In particular, the Lifshitz spacetime (which
would play the role of ground state in the thermodynamic description)\textcolor{black}{{}
suffer from divergent tidal forces. Additionally, asymptotically Lifshitz
black holes are not vacuum solutions to General Relativity, and it
is mandatory to include extra matter fields as in the case of Proca
fields \cite{Marika}, p-form gauge fields \cite{Kachru,Pang}, to
name a couple of examples.}

In the present work, we will adopt a different approach to this holographic
realization, where the anisotropic scaling properties of the boundary
field theory instead emerge from a very special choice of boundary
conditions for General Relativity on AdS$_{3}$. This new set of boundary
conditions is labeled by a nonnegative integer $n$, and is related
with the Korteweg-de Vries (KdV) hierarchy of integrable systems \cite{KdV}\footnote{Other examples of this relationship between 2D integrable systems
and gravity in 2+1, have been also made for the cases of ``flat''
and ``soft hairy'' boundary conditions in \cite{flat KdV} and \cite{Emilio},
respectively.}. Allowing that at the asymptotic region the Lagrange multipliers
could depend on the global charges, it was shown that the reduced
phase space of Einstein field equations is given precisely by two
copies of the $n$-th member of the KdV hierarchy. It is worth to
emphasize that, although these boundary conditions describe asymptotically
locally AdS$_{3}$ spacetimes, the associated dual field theory possesses
an anisotropic scaling of Lifshitz type, 
\begin{equation}
t\rightarrow\lambda^{z}t\,,\quad\phi\rightarrow\lambda\phi\,,
\end{equation}
where the dynamical exponent is given by $z=2n+1$. In the context
of black hole thermodynamics, KdV-type boundary conditions define
an unconventional thermodynamic ensemble, which leads to a generalized
thermodynamic description of the BTZ black hole. Remarkably, this
thermodynamic description shows very similar features with the ones
found in the study of Lifshitz black holes \cite{Bertoldi,Myung,Ay=00003D0000F3n-Beato,Brenna,Bravo Gaete},
and has a deep relationship with the work of Hardy and Ramanujan on
the counting of partitions of an integer into $z$-th powers \cite{Melnikov}.
In this work, we will focus on the main characteristics that this
thermodynamic ensemble implies, and its differences with the standard
analysis.

The paper is organized as follows. In Section \ref{Section II} we
provide a brief review of KdV-type boundary conditions in the context
of Chern-Simons description of General Relativity with negative cosmological
constant in three dimensions. In Section \ref{Section III} it is
shown that an anisotropic Smarr formula emerges from the radially
conserved charge associated with the scale invariance of the reduced
Einstein-Hilbert action endowed with KdV-type boundary conditions.
Section \ref{Section IV}, is devoted to the anisotropic Cardy formula
and the relation with its Smarr counterpart. Finally, the thermal
stability of BTZ black hole with KdV-type boundary conditions is deeply
analyzed in Section \ref{Section V}. We conclude with some comments
in Section \ref{Section VI}.

\section{General Relativity on AdS$_{3}$ and the KdV-type boundary conditions\label{Section II}}

General Relativity with negative cosmological constant in three dimensional
spacetimes can be formulated as the difference of two Chern-Simons
actions for gauge fields $A^{\pm}$, evaluated on two independent
copies of the $sl(2,\mathbb{R})$ algebra \cite{AchucarroTownsend,Witten},
\begin{equation}
I_{EH}=I_{CS}[A^{+}]-I_{CS}[A^{-}]\,,\label{EinsteinHilbert}
\end{equation}
The above action corresponds precisely to General Relativity on AdS$_{3}$
only if both Chern-Simons levels are given by $k=\ell/4G$, where
$\ell$ is the AdS radius and $G$ the Newton constant.

In order to describe the asymptotic form of the fields, it is convenient
to do the analysis using auxiliary fields $a^{\pm}$, which are defined
by a precise gauge transformation on $A^{\pm}$ \cite{Coussaert},
\begin{equation}
A^{\pm}=b_{\pm}^{-1}\left(a^{\pm}+d\right)b_{\pm}\,,
\end{equation}
with $b_{\pm}=e^{\pm\text{log}(r/\ell)L_{0}^{\pm}}$\footnote{The representation that we use is the same as in \cite{Generalized}.}.
So that, the radial component of $a^{\pm}$ vanishes, while the remain
ones only depend on time and the angular coordinate. Proceeding as
in \cite{Generalized,Chemical}, the non-vanishing components of the
auxiliary fields are given by 
\begin{equation}
a_{\phi}^{\pm}=L_{\pm}-\frac{1}{4}\mathcal{L}_{\pm}L_{\mp}\,,\quad a_{t}^{\pm}=\pm\mu_{\pm}L_{\pm}-\partial_{\phi}\mu_{\pm}L_{0}\pm\frac{1}{2}\left(\partial_{\phi}^{2}\mu_{\pm}-\frac{1}{2}\mu_{\pm}\mathcal{L}_{\pm}\right)L_{\mp}\,,\label{gauge fields}
\end{equation}
where $\mathcal{L}_{\pm}(t,\phi)$ stand for the dynamical fields,
and $\mu_{\pm}(t,\phi)$ correspond to the values of the Lagrange
multipliers at infinity. In the asymptotic region, the field equations
reduce to 
\begin{equation}
\partial_{t}\mathcal{L}_{\pm}=\pm\mathcal{D}^{\pm}\mu_{\pm}\,,\quad\mathcal{D}^{\pm}:=(\partial_{\phi}\mathcal{L}_{\pm})+2\mathcal{L}_{\pm}\partial_{\phi}-2\partial_{\phi}^{3}\,.\label{L punto}
\end{equation}

It is worth highlighting that the boundary conditions are fully specified
once a precise form of the Lagrange multipliers at infinity is provided.
In the standard approach of Brown and Henneaux \cite{BrownHenneaux},
the Lagrange multipliers are set as $\mu_{\pm}=1$. Going one step
further, one can generalize this analysis by choosing arbitrary functions
of the coordinates, $\mu_{\pm}=\mu_{\pm}(t,\phi)$, which, in order
to have a well-defined action principle, are held fixed at the boundary
($\delta\mu_{\pm}=0$) \cite{Generalized,Chemical}. However, even
beyond that, one can still guarantees the integrability of the boundary
term in the action, if one allow that the Lagrange multipliers may
depend on the dynamical fields and their spatial derivatives, giving
rise to a complete new set of boundary conditions. Here, we will focus
on the family of KdV-type boundary conditions, introduced in \cite{KdV},
which are labeled by a non negative integer $n$. In this context,
the Lagrange multipliers are chosen to be given by the $n$-th Gelfand-Dikii
polynomial evaluated on $\mathcal{L}_{\pm}$ and can be obtained by
the functional derivative with respect to $\mathcal{L}_{\pm}$ of
the $n$-th Hamiltonian of the KdV hierarchy, i.e., 
\begin{equation}
\mu_{\pm}^{(n)}[\mathcal{L}_{\pm}]=\frac{\delta H_{\pm}^{(n)}}{\delta\mathcal{L}_{\pm}}\,,\label{mu}
\end{equation}
where the following recursion relation is satisfied 
\begin{equation}
\partial_{\phi}\mu_{\pm}^{(n+1)}=\frac{n+1}{2n+1}\mathcal{D}^{\pm}\mu_{\pm}^{(n)}\,.\label{Recurrencia}
\end{equation}
Thus, for the case $n=0$, one recovers the Brown-Henneaux boundary
conditions ($\mu_{\pm}^{(0)}=1$), and in consequence, according to
(\ref{L punto}), the dynamical fields are chiral. In the case $n=1$,
the Lagrange multipliers are given by $\mu_{\pm}^{(1)}=\mathcal{L}_{\pm}$,
and then the field equations reduce to two copies of the KdV equation,
while for the remaining cases ($n>1$) the field equations are given
by the corresponding $n$-th member of the KdV hierarchy.

As a consequence of the reduction of Einstein field equations to the
KdV hierarchy\footnote{As shown in \cite{Pinox=00003D000026Matulich}, by performing the
Hamiltonian reduction of KdV-type boundary conditions, the equations
\eqref{L punto} actually corresponds to the conservation law of the
energy-momentum tensor of the corresponding theory at the boundary.
For the particular case $n=0$, the field equations are equivalent
to the aforementioned conservation law. See also \cite{Grumiller Wout},
for a recent related result, in the case of ``near horizon'' boundary
conditions.}, the ``boundary gravitons'' (global gravitational excitations)
possess an anisotropic scaling of Lifshitz type 
\begin{equation}
t\rightarrow\lambda^{z}t\,,\quad\phi\rightarrow\lambda\phi\,,\quad\mathcal{L}_{\pm}\rightarrow\lambda^{-2}\mathcal{L}_{\pm}\,,\label{KdV scaling}
\end{equation}
where the dynamical exponent $z$ is related to the KdV label $n$
by $z=2n+1$. It must be remarked that, although the solutions of
Einstein field equations are locally AdS$_{3}$, they inherit an anisotropic
scaling from the choice of KdV-type boundary conditions. For this
reason, the thermodynamic properties of black holes will also carry
a $z$-dependence.

According to the canonical approach \cite{ReggeTeitelboim}, the variation
of the generators of the asymptotic symmetries are readily found to
be given by $\delta Q=\delta Q_{+}[\varepsilon_{+}]-\delta Q_{-}[\varepsilon_{-}]$,
where 
\begin{equation}
\delta Q_{\pm}[\varepsilon_{\pm}]=-\frac{\ell}{32\pi G}\int d\phi\,\varepsilon_{\pm}\delta\mathcal{L}_{\pm}\,.\label{delta Q}
\end{equation}

In particular, when the gauge parameters are related with the asymptotic
Killing vectors $\partial_{\phi}$ and $\partial_{t}$, one can integrate
\eqref{delta Q} directly. Indeed, the angular momentum is given by
\begin{equation}
Q\left[\partial_{\phi}\right]=\frac{\ell}{32\pi G}\int d\phi\,\left(\mathcal{L}_{+}-\mathcal{L}_{-}\right)\,,
\end{equation}
while for time translations, 
\begin{equation}
\delta Q\left[\partial_{t}\right]=\frac{\ell}{32\pi G}\int d\phi\left(\mu_{+}\delta\mathcal{L}_{+}+\mu_{-}\delta\mathcal{L}_{-}\right)\,,
\end{equation}
which by virtue of \eqref{mu}, the energy integrate as 
\begin{equation}
Q\left[\partial_{t}\right]=\frac{\ell}{32\pi G}\left(H_{+}^{(k)}+H_{-}^{(k)}\right)\,.\label{energy}
\end{equation}

\subsection{The BTZ black hole with KdV-type boundary conditions}

For each allowed choice of $n$ (or equivalently $z$), the spectrum
of solutions is quite different. Nonetheless, BTZ black hole \cite{BTZ,BHTZ}
fits within every choice of boundary conditions in (\ref{mu}). Indeed,
this class of configurations is described by constant $\mathcal{L}_{\pm}$,
which trivially solves (\ref{L punto}) for all possible values of
$n$. In this case, according to the normalization choice in \eqref{Recurrencia},
it is possible to show that the Lagrange multipliers generically acquire
a remarkably simple form, $\mu_{\pm}^{(n)}=\mathcal{L}_{\pm}^{n}N_{\pm}$,
where $N_{\pm}$ is assumed to be fixed without variation at the boundary
($\delta N_{\pm}=0$). Note that $\mu_{\pm}^{(0)}=N_{\pm}$, so in
that special case, the Lagrange multipliers at infinity are held constants
but given by arbitrary values, and the standard Brown-Henneaux analysis
\cite{BrownHenneaux} is recovered by setting $N_{\pm}=1$. In this
scenario, along the lines of \cite{Chemical}, the Lagrange multipliers
are allowed to depend on the dynamical fields, which amounts to a
different fixing of the ``chemical potentials'' at the boundary,
implying that we are dealing with the same black hole configuration
but in a different thermodynamic ensemble (see e.g., \cite{Grumiller}).
In what follows, we will use the dynamical exponent $z$, instead
of the KdV-label $n$, so by using $z=2n+1$, we can rewrite the KdV-type
Lagrange multipliers as\footnote{In the context of AdS/CFT holography, the relationship between the
chemical potentials and conserved charges is known as ``multi-trace
deformations'' of the dual theory \cite{Generalized Gibbs 1,Generalized Gibbs 2,Generalized Gibbs 3,Generalized Gibbs 4}.}, 
\begin{equation}
\mu_{\pm}=\mathcal{L}_{\pm}^{\frac{z-1}{2}}N_{\pm}\;.\label{Anisotropic mu}
\end{equation}

The energies of the left and right movers also takes a simple form
for a generic choice of $n$, namely $E_{\pm}=\frac{\ell}{32\pi G}H_{\pm}^{(n)}=\frac{\ell}{16G}\frac{1}{n+1}\mathcal{L}_{\pm}^{n+1}$.
Therefore, in terms of the dynamical exponent we can rewrite them
as 
\begin{equation}
E_{\pm}=\frac{\ell}{8G}\frac{1}{z+1}\mathcal{L}_{\pm}^{\frac{z+1}{2}}\,.\label{Anisotropic E}
\end{equation}
From the gravitational perspective, according to \eqref{energy},
the energy of the BTZ black hole is determined by $E=E_{+}+E_{-}\,.$

In the next section we will show that an anisotropic version of Smarr
formula naturally emerges as the consequence of the scale invariance
of the reduced Einstein-Hilbert action, as long as we consider the
KdV-type boundary conditions.

\section{The anisotropic Smarr formula as a radial conservation law\label{Section III}}

In \cite{BanadosTheisen}, the authors showed that the reduced Einstein-Hilbert
action coupled to a scalar field in AdS$_{3}$ spacetimes is invariant
under a set of scale transformations which leads to a radial conservation
law by using the Noether theorem. When this conserved quantity is
evaluated in a particular solution of the theory, namely, a black
hole solution, one obtains a Smarr relation \cite{SMARR}. This method
has been successfully applied to several cases in the literature \cite{Hyun,Ahn,Charged Smarr-EFR,Bravo-Gaete,Gonzalez,Ahn-1,Wu,Kim}
for different theories. By following this procedure, we show that
it is possible as well, to obtain a generalization of the Smarr formula
for the BTZ black hole endowed with KdV-type boundary conditions.
As a consequence, the entropy as a bilinear form of the global charges
of the black hole, manifestly depends on the dynamical exponent.

In the metric formulation, the Einstein-Hilbert action has the following
form

\begin{equation}
I_{EH}=\int d^{3}x\sqrt{-g}\left[\frac{1}{2\kappa}\left(R-2\Lambda\right)\right]\,,\label{EH action}
\end{equation}
where $\kappa=8\pi G$, and the cosmological constant is related to
the AdS radius by $\Lambda=-\ell^{-2}$.

By considering stationary and circularly symmetric spacetimes described
by the following line element 
\begin{equation}
ds^{2}=-\mathcal{N}\left(r\right)^{2}\mathcal{F}\left(r\right)^{2}dt^{2}+\frac{dr^{2}}{\mathcal{F}\left(r\right)^{2}}+r^{2}\left(d\phi+\mathcal{N}^{\phi}\left(r\right)dt\right)^{2}\,,\label{Stationary metric}
\end{equation}
the reduced action principle in the canonical form is given by 
\begin{equation}
I=-2\pi\left(t_{2}-t_{1}\right)\int dr\left(\mathcal{NH}+\mathcal{N^{\phi}}\mathcal{H}_{\phi}\right)+B\,,\label{reduced action}
\end{equation}
where the boundary term $B$ must be added in order to have a well-defined
variational principle. The surface deformation generators $\mathcal{H}$,
$\mathcal{H_{\phi}}$ acquire the following form 
\begin{align}
\mathcal{H} & =-\frac{r}{\kappa\ell^{2}}+4\kappa r(\pi^{r\phi})^{2}+\frac{\left(\mathcal{F}^{2}\right)^{\prime}}{2\kappa}\,,\label{H}\\
\mathcal{H}_{\phi} & =-2(r^{2}\pi^{r\phi})^{\prime}\,,\label{Hphi}
\end{align}
where $\mathcal{N}$ and $\mathcal{N}^{\phi}$ stand for their corresponding
Lagrange multipliers. The only nonvanishing component of the momenta
$\pi^{ij}$ is explicitly given by 
\begin{equation}
\pi^{r\phi}=-\frac{\left(\mathcal{N}^{\phi}\right)^{\prime}r}{4\kappa\mathcal{N}}\,,\label{pr}
\end{equation}
where prime denotes derivative with respect to $r$.

The above reduced action principle turns out to be invariant under
the following set of scale transformations 
\begin{equation}
\bar{r}=\xi r\,,\quad\bar{\mathcal{N}}=\xi^{-2}\mathcal{N}\,,\quad\bar{\mathcal{N^{\phi}}}=\xi^{-2}\mathcal{N^{\phi}}\,,\quad\bar{\mathcal{F}^{2}}=\xi^{2}\mathcal{F}^{2}\,,\label{radial scaling}
\end{equation}
where $\xi$ is a positive constant. By applying the Noether theorem,
we obtain a radially conserved charge associated with the aforementioned
symmetries, 
\begin{equation}
C(r)=\frac{1}{4G}\left[-\mathcal{N}\mathcal{F}^{2}+\frac{r\mathcal{N}\left(\mathcal{F}^{2}\right)^{\prime}}{2}-\frac{r^{3}\left(\mathcal{N}^{\phi}\right)^{\prime}\mathcal{N}^{\phi}}{\mathcal{N}}\right]\,,
\end{equation}
which means that $C'=0$ on-shell.

We will find a Smarr formula by exploiting the fact that this conserved
charge must satisfy $C(\infty)=C(r_{+})$, where $r_{+}$ is the event
horizon of the BTZ black hole solution with KdV-type boundary conditions.

\subsection{Conserved charge at infinity}

For the class of configurations considered here, the metric functions
in ``Schwarzschild'' coordinates are given by 
\begin{eqnarray}
\mathcal{N}(r) & = & \frac{\ell}{2}\left(\mu_{+}+\mu_{-}\right)\,,\nonumber \\
\mathcal{N}^{\phi}(r) & = & \frac{1}{2}\left(\mu_{+}-\mu_{-}\right)+\frac{\ell^{2}}{8r^{2}}\left(\mathcal{L}_{+}-\mathcal{L}_{-}\right)\left(\mu_{+}+\mu_{-}\right)\,,\label{metric functions}\\
\mathcal{F}^{2}(r) & = & \frac{r^{2}}{\ell^{2}}-\frac{1}{2}\left(\mathcal{L}_{+}+\mathcal{L}_{-}\right)+\frac{\ell^{2}}{16r^{2}}\left(\mathcal{L}_{+}-\mathcal{L}_{-}\right)^{2}\,.\nonumber 
\end{eqnarray}
where $\mu_{\pm}$ corresponds to the arbitrary values of the Lagrange
multipliers at infinity, leading to a simple expression for the radially
conserved charge at infinity, 
\begin{equation}
C(\infty)=\frac{\ell}{8G}\left(\mu_{+}\mathcal{L}_{+}+\mu_{-}\mathcal{L}_{-}\right)\,.
\end{equation}
Thus in the case of KdV-type boundary conditions \eqref{Anisotropic mu},
can be written as

\begin{equation}
C(\infty)=\frac{\ell}{8G}\left(N_{+}\mathcal{L}_{+}^{\frac{z+1}{2}}+N_{-}\mathcal{L}_{-}^{\frac{z+1}{2}}\right)\,,
\end{equation}
which in terms of the left and right energies \eqref{Anisotropic E},
reads 
\begin{equation}
C(\infty)=(z+1)N_{+}E_{+}+(z+1)N_{-}E_{-}\,.
\end{equation}

\subsection{Conserved charge at the event horizon}

To evaluate the radial charge at the event horizon we must ensure
that the Euclidean configuration is smooth around this point. The
inner and outer horizons $r_{\pm}$ in these coordinates, are determined
by $\mathcal{F}^{2}(r_{\pm})=0$, where $r_{\pm}=\frac{\ell}{2}\left(\sqrt{\mathcal{L}_{+}}\pm\sqrt{\mathcal{L}_{-}}\right)$.
In consequence, it is clear that the regularity requirement in this
gauge translates into, 
\begin{equation}
\mathcal{N}(r_{+})\mathcal{F}^{2}(r_{+})^{\prime}=4\pi\,,\quad\mathcal{N}^{\phi}(r_{+})=0\,,\label{rc}
\end{equation}
which implies that the Euclidean metric becomes regular for $\mu_{\pm}=\frac{2\pi}{\sqrt{\mathcal{L}_{\pm}}}$.
Considering the regularity conditions \eqref{rc} and the metric functions
$\mathcal{F}^{2}$, $\mathcal{N}$ and $\mathcal{N}^{\phi}$, the
value of the radial charge at the event horizon is 
\begin{equation}
C(r_{+})=\frac{\pi r_{+}}{2G}=S\,,
\end{equation}
which corresponds to the entropy of the BTZ black hole.

Now, by making use of the equality $C(r_{+})=C(\infty)$, the anisotropic
Smarr formula is obtained\footnote{Resembling expressions for the entropy as a bilinear combination of
the global charges times the chemical potentials have been previously
found for three dimensional black holes and cosmological configurations
in the context of higher spin gravity \cite{Generalized,Gary,Matulich},
hypergravity \cite{Henneaux,Henneaux-1} and extended supergravity
\cite{Fuentealba}. The factor in front of each term corresponds to
the conformal weight (spin) of the corresponding generator. }, 
\begin{equation}
S=(z+1)N_{+}E_{+}+(z+1)N_{-}E_{-}\,.\label{Anisotropic Smarr}
\end{equation}

Identifying the Lagrange multipliers as the inverse of left and right
temperatures $T_{\pm}=N_{\pm}^{-1}$, the above expression acquires
the following form, 
\begin{equation}
S=(z+1)\frac{E_{+}}{T_{+}}+(z+1)\frac{E_{-}}{T_{-}}\,.\label{Anisotropic Smarr-1}
\end{equation}

Turning off the angular momentum, this expression reduces to\footnote{The left and right temperatures are related with the Hawking temperature
through $T=\frac{2T_{+}T_{-}}{\left(T_{+}+T_{-}\right)}$, so in the
absence of rotation $T_{+}=T_{-}=T$.} 
\begin{equation}
E=\frac{1}{(z+1)}TS\,,
\end{equation}
which is in agreement with Smarr formula for Lifshitz black holes
in three dimensions found in the literature (see e.g. \cite{Ay=00003D0000F3n-Beato,Brenna}).
It is worth to point out that since the BTZ black hole with generic
boundary conditions is asymptotically AdS$_{3}$, the contribution
due to the rotation naturally appears in the anisotropic Smarr formula
\eqref{Anisotropic Smarr}, despite the fact that, as far as the knowledge
of the present authors, there is no a rotating Lifshitz black hole
in three dimensions. It is also worth mentioning that in the limit
$z\rightarrow0$, \eqref{Anisotropic Smarr} fits with the corresponding
Smarr relation of soft hairy horizons in three spacetimes dimensions
\cite{Afshar}.

It is worth to remark that the scaling \eqref{radial scaling} is
equivalent to the scaling of the Lifshitz type introduced in \eqref{KdV scaling}.
By redefining the scale factor as $\lambda\rightarrow\xi^{-1}$ in
\eqref{radial scaling}, we obtain 
\begin{equation}
t\rightarrow\xi^{-z}t\,,\quad r\rightarrow\xi r\,,\quad\phi\rightarrow\xi^{-1}\phi\,,\quad\mathcal{L}_{\pm}\rightarrow\xi^{2}\mathcal{L}_{\pm}\,.
\end{equation}
Then, in the case of KdV-type boundary conditions \eqref{Anisotropic mu},
we can see that the Lagrange multipliers at infinity scales as $\bar{\mu}_{\pm}=\xi^{z-1}\mu_{\pm}$,
therefore from \eqref{metric functions}, we can deduce that $\mathcal{N}$
and $\mathcal{N^{\phi}}$ scales accordingly. Now, since the reduced
Hamiltonian action does not depend on $t$ and $\phi$, before integrate
them, we can absorb its scalings on the Lagrange multipliers, and
in consequence, they must scale as $\bar{\mathcal{N}}=\xi^{-2}\mathcal{N}$
and $\bar{\mathcal{N}}^{\phi}=\xi^{-2}\mathcal{N^{\phi}}$, which
is in full agreement with \eqref{radial scaling}.

In the following chapter, we will show that by considering a two dimensional
field theory defined on the torus, it is possible to recover the anisotropic
Smarr formula by means of an anisotropic version of the standard S-modular
invariance of the partition function.

\section{The anisotropic Cardy formula\label{Section IV}}

On this section, we will tackle the thermodynamical description of
the BTZ black hole from an holographic perspective. The partition
function of the 2D dual field theory is made up by the contribution
of two non interacting left and right systems, each one with a corresponding
temperature given by $T_{+}=\beta^{-1}$and $T_{-}=\beta_{-}^{-1}$.
The modular parameter of the torus $\tau$, where the theory is defined,
is related to the left and right periods of the thermal cycle, $\beta_{+}$
and $\beta_{-}$, by 
\begin{equation}
\tau=\frac{i\beta_{+}}{2\pi}\ ,\qquad\bar{\tau}=-\frac{i\beta_{-}}{2\pi}\ .
\end{equation}

As argued in \cite{KdV} and \cite{GTT}, the partition function of
the dual field theory at the boundary is invariant under the anisotropic
S-duality transformation, given by 
\begin{equation}
Z[\beta_{\pm};z]=Z[(2\pi)^{1+\frac{1}{z}}\beta_{\pm}^{-\frac{1}{z}};z^{-1}]\,,\label{Zlow-high}
\end{equation}
and, by assuming the existence of a gap in the energy spectrum it
is possible to write the partition function as the contribution of
two pieces 
\begin{equation}
Z(\beta_{+},\beta_{-})=e^{-I(\beta_{+},\beta_{-})}+e^{-I_{\text{0}}(\beta_{+},\beta_{-})}\,,
\end{equation}
such that, at low temperatures, the contribution of the ground state
$I_{\text{0}}$ dominates the partition function, so it can be approximated
by 
\begin{equation}
Z(\beta_{+},\beta_{-})\approx\exp\left(-\beta_{+}E_{+}^{0}-\beta_{-}E_{-}^{0}\right)\,,
\end{equation}
and by virtue of the anisotropic S-duality, at high temperature regime
the partition function acquires the following form 
\begin{equation}
Z(\beta_{+},\beta_{-})\approx\exp\left(-\frac{(2\pi)^{1+\frac{1}{z}}}{\beta_{+}^{\frac{1}{z}}}E_{+}^{0}[z^{-1}]-\frac{(2\pi)^{1+\frac{1}{z}}}{\beta_{-}^{\frac{1}{z}}}E_{-}^{0}[z^{-1}]\right)\,.\label{Partition-1}
\end{equation}

It is well known that by taking the inverse Laplace transform of the
partition function \eqref{Partition-1} we can obtain the asymptotic
growth of the density of states 
\begin{equation}
\begin{aligned}\rho(E_{+},E_{-}) & =\frac{1}{\left(2\pi i\right)^{2}}\int_{-i\infty}^{+i\infty}d\beta_{+}d\beta_{-}e^{\beta_{+}E_{+}+\beta_{-}E_{-}}Z(\beta_{+},\beta_{-})\\
 & \approx\frac{1}{\left(2\pi i\right)^{2}}\int d\beta_{+}d\beta_{-}e^{f_{+}+f_{-}}\,,
\end{aligned}
\label{DensityStates}
\end{equation}
where the function $f_{\pm}$ is defined as 
\begin{equation}
f_{\pm}(\beta_{\pm},E_{\pm},E_{\pm}^{0}):=-\frac{(2\pi)^{1+\frac{1}{z}}}{\beta_{\pm}^{\frac{1}{z}}}E_{\pm}^{0}[z^{-1}]+\beta_{\pm}E_{\pm}\,.
\end{equation}
We can evaluate this expression using the saddle-point approximation
for fixed energies in the limit of $E_{\pm}\gg\vert E_{\pm}^{0}\vert$.
Indeed, there is a critical point $\beta_{\pm}(E_{\pm},E_{\pm}^{0})$,
\begin{equation}
\beta_{\pm}=2\pi\left(\frac{-E_{\pm}^{0}[z^{-1}]}{zE_{\pm}}\right)^{\frac{z}{1+z}}\,.\label{CriticalPoint}
\end{equation}
Then, the entropy of the system 
\begin{equation}
S=\log\rho\approx f_{+}(\beta_{+},E_{+},E_{+}^{0})+f_{-}(\beta_{-},E_{-},E_{-}^{0})\;,\label{LogS}
\end{equation}
will be given by two copies of the anisotropic Cardy formula\footnote{As explained in \cite{KdV}, for odd values of $n=(z-1)/2$, Euclidean
BTZ with KdV-type boundary conditions is diffeomorphic to thermal
AdS$_{3}$, but with reversed orientation, and in consequence, there
is a opposite sign between Euclidean and Lorentzian left and right
energies of the ground state. As it will be shown in the next section,
this leads to a local thermodynamic instability of the system for
odd values of $n$. So, it is mandatory to adopt $E_{\pm}^{0}\left[z^{-1}\right]\rightarrow-\left|E_{\pm}^{0}\left[z^{-1}\right]\right|$,
in the Lorentzian ground state energies of the anisotropic Cardy formula.}\emph{ }\cite{KdV}, \cite{GTT}, 
\begin{equation}
S=2\pi\left(z+1\right)\left[\left(\frac{\vert E_{+}^{0}[z^{-1}]\vert}{z}\right)^{z}E_{+}\right]^{\frac{1}{z+1}}+2\pi\left(z+1\right)\left[\left(\frac{\vert E_{-}^{0}[z^{-1}]\vert}{z}\right)^{z}E_{-}\right]^{\frac{1}{z+1}}\,.\label{CardyLKdV-1}
\end{equation}

Remarkably, if instead, we consider the ground state energies in terms
of the inverse temperatures and the left and right energies, 
\begin{equation}
\left|E_{\pm}^{0}[z^{-1}]\right|=zE_{\pm}[z]\left(\frac{\beta_{\pm}}{2\pi}\right)^{1+\frac{1}{z}}\,,\label{CriticalPoint-1}
\end{equation}
in \eqref{LogS}, we found that the entropy reduces to 
\[
S=(z+1)E_{+}\beta_{+}+(z+1)E_{-}\beta_{-}\,,
\]
which exactly matches with the anisotropic Smarr formula previously
obtained by considering the scale symmetry of the Einstein-Hilbert
reduced action \eqref{Anisotropic Smarr-1}. This relationship between
Cardy and Smarr formulas has been previously suggested in the literature
\cite{Ay=00003D0000F3n-Beato}. Interestingly enough, the link between
both expressions is the anisotropic version of the Stefan-Boltzmann
law, which is nothing else than the relation between energy and temperature
given by the critical point \eqref{CriticalPoint-1}.

From equation \eqref{CardyLKdV-1}, it is clear that the entropy written
as a function of left and right energies scales as 
\begin{equation}
\begin{aligned}S\left(\lambda E_{+},\lambda E_{-}\right) & =\lambda^{\sigma}S\left(E_{+},E_{-}\right)\end{aligned}
\,,
\end{equation}
where 
\begin{equation}
\sigma=\frac{1}{\left(z+1\right)}\,.
\end{equation}
Hence, it is reassuring to prove that, by simply applying the Euler
theorem for homogeneous functions, gives 
\begin{equation}
S\left(E_{+},E_{-}\right)=\left(z+1\right)\left(E_{+}\beta_{+}+E_{-}\beta_{-}\right)\,,
\end{equation}
in the same way than the original derivation found in \cite{SMARR}.

The following section is devoted to the local and global thermal stability
of the BTZ black hole endowed with KdV-type boundary conditions.

\section{Thermodynamic stability and phase transitions\label{Section V}}

We analyze the thermodynamic stability at fixed chemical potentials.
Local stability condition can be determined by demanding a negative
defined Hessian matrix of the free energy of the system (see e.g.
\cite{Celine}). Nonetheless, in this ensemble it can equivalently
be performed by the analysis of the left and right specific heats
with fixed chemical potential. From the anisotropic Stefan-Boltzmann
law 
\begin{equation}
E_{\pm}\left[z\right]=\frac{1}{z}\vert E_{\pm}^{0}\left[z^{-1}\right]\vert\left(2\pi\right)^{1+\frac{1}{z}}T_{\pm}^{1+\frac{1}{z}}\,,\label{StefanBoltzmann}
\end{equation}
one finds that left and right specific heats are given by

\begin{equation}
C_{\pm}\left[z\right]=\frac{\partial E_{\pm}}{\partial T_{\pm}}=\frac{z+1}{z^{2}}\vert E_{\pm}^{0}\left[z^{-1}\right]\vert(2\pi)^{1+\frac{1}{z}}T_{\pm}^{\frac{1}{z}}\,.
\end{equation}
We see that, for all possible values of $z$, the specific heats are
continuous monotonically increasing functions of $T_{\pm}$, and always
positive\footnote{Strictly speaking, specific heats $C_{\pm}$ are always positive provided
that $T_{\pm}>0$. In terms of the temperature and angular velocity
of the black hole, the above is equivalent to the non-extremality
condition; $0<T\,,\ -1<\Omega<1\,$. Since in the present paper we
are not dealing with the extremal case, we will consider that this
condition is always fulfilled.}, which means that the system is at least locally stable. It is important
to remark that, as mentioned at the end of the last section, if we
had not warned on the correct sign of the ground state energies for
odd $n$, the sign of the specific heats would have depended on $z$,
and in consequence, for odd values of $n$ the black hole would be
thermodynamically unstable.

Since the specific heats are finite and positive regardless of the
value of $z$, the BTZ black hole with generic KdV-type boundary conditions
can always reach local thermal equilibrium with the heat bath at any
temperature.

Once local stability is assured, it makes sense to ask about the global
stability of the system. Following the seminal paper of Hawking and
Page \cite{HawkingPage}, we use the free energies at fixed values
of the chemical potentials of the two phases present in the spectrum
(BTZ and thermal AdS$_{3}$), in order to realize which one is thermodynamically
preferred.

In the semiclassical approximation, the on-shell Euclidean action
is proportional to the free energy of the system. Taking into account
the contributions of the left and right movers, the action acquires
the following form 
\begin{equation}
I=N_{+}E_{+}+N_{-}E_{-}-S\,.\label{L and R ensemble}
\end{equation}
Assuming a non-degenerate ground state with zero entropy and whose
left and right energies are equal and negative defined, $E_{\pm}\rightarrow-\left|E^{0}\left[z\right]\right|$,
we see that the value of the action of the ground state is given by
\begin{equation}
I_{0}=-\vert E^{0}\left[z\right]\vert\left(\frac{1}{T_{+}}+\frac{1}{T_{-}}\right)\;.\label{AdS action}
\end{equation}
On the other hand, considering in \eqref{L and R ensemble} a system
whose entropy is given by the formula \eqref{Anisotropic Smarr},
we obtain that 
\begin{equation}
I=-z\left(N_{+}E_{+}+N_{-}E_{-}\right)\,,\label{I}
\end{equation}
hence, using the formulae for the energies in \eqref{StefanBoltzmann},
the action then reads 
\begin{equation}
I=-\vert E^{0}\left[z^{-1}\right]\vert\left(2\pi\right)^{\frac{z+1}{z}}\left(T_{+}^{\frac{1}{z}}+T_{-}^{\frac{1}{z}}\right)\,.\label{BTZ action}
\end{equation}
Therefore, it is straightforward to see that, regardless of the value
of $z$, the partition function $Z=e^{I+I_{0}}$, will be dominated
by \eqref{AdS action} at low temperatures, and by \eqref{BTZ action}
at the high temperatures regime. It can also be shown that, consistently,
we are able to found the same ground state action by making use of
the anisotropic S-duality transformation \eqref{Zlow-high} on \eqref{BTZ action}.

In what follows, we will focus on the simplest case where the whole
system is in equilibrium at a fixed temperature $T_{\pm}=T$. Then,
the free energy of the system at high and low temperatures will respectively
given by 
\begin{equation}
F=-2\vert E^{0}\left[z^{-1}\right]\vert\left(2\pi\right)^{1+\frac{1}{z}}T^{1+\frac{1}{z}}\,,\qquad F_{0}=-2\vert E^{0}\left[z\right]\vert\,,
\end{equation}
and comparing them, we can obtain the self-dual temperature, at where
both free energies coincide, 
\begin{equation}
T_{s}\left[z\right]=\frac{1}{2\pi}\left\vert \frac{E^{0}\left[z\right]}{E^{0}\left[z^{-1}\right]}\right\vert ^{\frac{z}{z+1}}\;,\label{T self-dual}
\end{equation}
which manifestly depend on the dynamical exponent. An interesting
remark is worth to be mentioned. The fact that the self-dual temperature
$T_{s}$ depends on the specific choice of boundary conditions, is
because the S-duality transformation involves an inversion of the
dynamical exponent between the high and low temperature regimes, namely,
$z\rightarrow z^{-1}$. This is a highly non trivial detail in the
calculation. If one does not take it into account, the self-dual temperature
would be the same for all values of $z$.

On the other hand, computing the free energies of the BTZ black hole
and thermal AdS$_{3}$ spacetime, we obtain 
\begin{equation}
F_{BTZ}=-\frac{\ell}{4G}\frac{z}{z+1}\left(2\pi T\right)^{\frac{z+1}{z}}\,,\qquad F_{AdS}=-\frac{\ell}{4G}\frac{1}{z+1}\,,
\end{equation}
and then, the self-dual temperature for which the two phases are equally
likely is 
\begin{equation}
T_{s}\left[z\right]=\frac{1}{2\pi}\left(\frac{1}{z}\right)^{\frac{z}{z+1}}\,.
\end{equation}
which exactly matches with \eqref{T self-dual}, if one identifies
the ground state energy of the field theory with the one of the AdS$_{3}$
spacetime with KdV-type boundary conditions, i.e., $E^{0}\rightarrow\frac{1}{2}E_{AdS}\left[z\right]=-\frac{\ell}{8G}\frac{1}{z+1}$.
\begin{figure}
\centering

\includegraphics[scale=0.5]{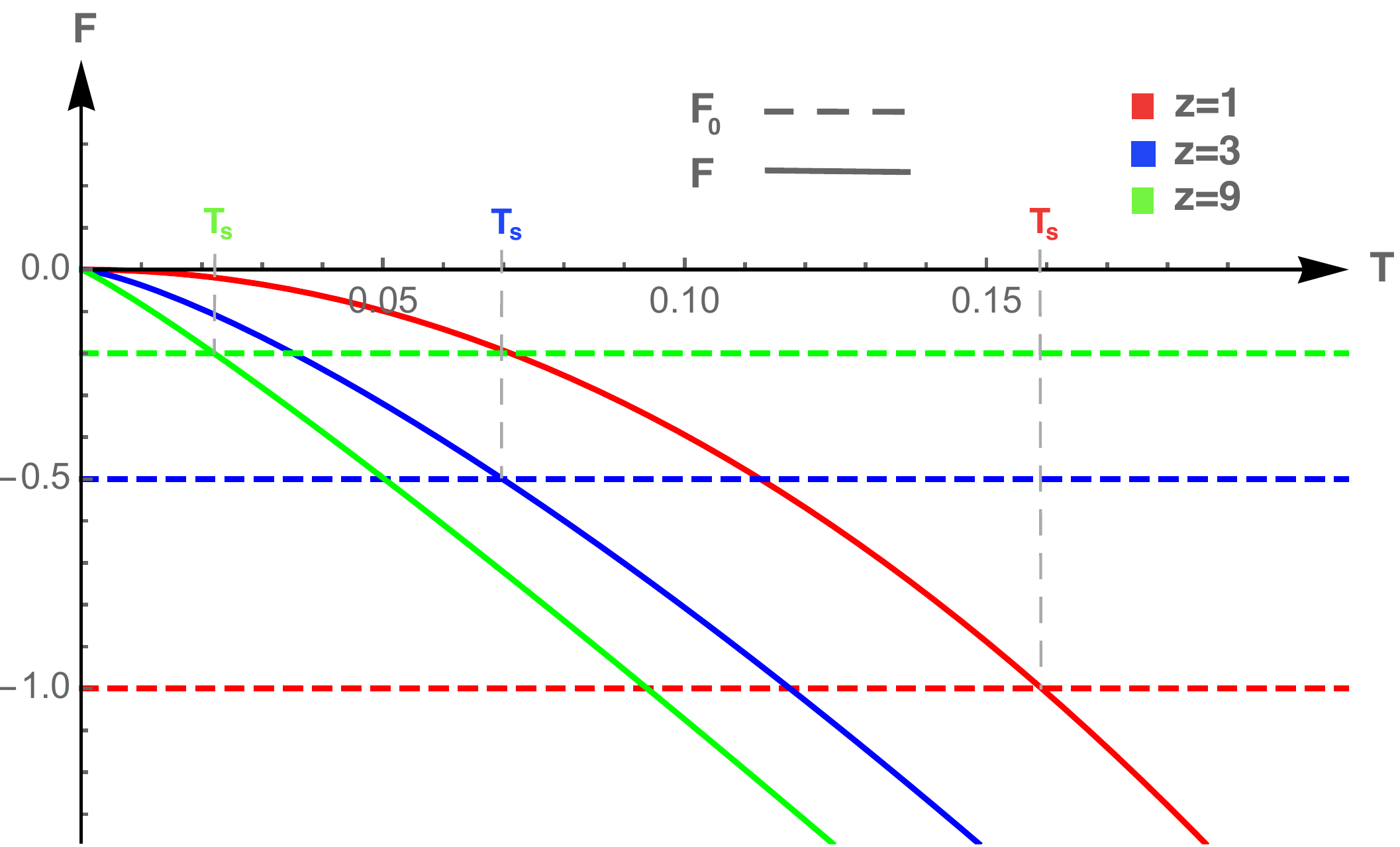}

\caption{Free energies of the BTZ black hole and thermal AdS$_{3}$ spacetime
with KdV-type boundary conditions associated to the dynamical exponents
$z=1\;;3\;;9$.}
\end{figure}

As it shown in Figure 1, for an arbitrary temperature below the self-dual
temperature ($T<T_{s}$), the thermal AdS$_{3}$ phase has less free
energy than the BTZ, and therefore the former one is the most probable
configuration, while if $T>T_{s}$, the black hole phase dominates
the partition function and hence is the preferred one. Note that for
higher values of $z$, the self-dual temperature becomes lower. The
latter point entails to a remarkable result. In the case of Brown-Henneaux
boundary conditions ($z=1$), one can deduce that in order for the
black hole reach the equilibrium with a thermal bath at the self-dual
point $T_{s}$, the event horizon must be of the size of the AdS$_{3}$
radius, i.e., $r_{+}=\ell$. Nonetheless, for a generic choice of
$z$, the horizon size has to be 
\begin{equation}
r_{+}^{s}\left[z\right]=\ell\left(\frac{1}{z}\right)^{\frac{1}{z+1}}\,.
\end{equation}
This means that the size of the black hole at the self-dual temperature
decreases for higher values of $z$. In the same way, at $T_{s}$,
the energy of the BTZ, $E_{BTZ}=E_{+}+E_{-}$, endowed with generic
KdV-type boundary conditions, acquires the following form 
\begin{equation}
E_{s}\left[z\right]=\frac{\ell}{4G}\frac{1}{z\left(z+1\right)}\,,
\end{equation}
and when compared to the AdS$_{3}$ spacetime energy, 
\begin{equation}
\Delta E=E_{BTZ}-E_{AdS}=\frac{\ell}{4G}\frac{1}{z}\,,
\end{equation}
we can observe that at the self-dual temperature there is an endothermic
process, where the system absorbs energy from the surround thermal
bath at a lower rate for higher values \LyXZeroWidthSpace \LyXZeroWidthSpace of
$z$.

From these last points we can conclude that the global stability of
the system is certainly sensitive to which KdV-type boundary condition
is chosen, since the free energy of the possible phases of the system
are explicitly $z$-dependent. Moreover, the temperature at which
both phases have equal free energy, the size of the black hole horizon
and the internal energy of the system at that temperature, decrease
for higher values of $z$, giving rise to a qualitatively different
behavior of the thermodynamic stability of the system, compared to
the standard analysis defined by $z=1$.

\section{Outlook and ending remarks\label{Section VI}}

The purpose of this work is twofold. On one hand, we have shown that
the anisotropic Smarr relation for the BTZ black hole endowed with
KdV-type boundary conditions can be obtained by following three different
approaches. First, by means of the Noether theorem, we obtained a
radial conserved quantity, which once evaluated in the BTZ solution
naturally leads to an expression for the entropy as a $z$-dependent
bilinear combination of the conserved charges times the chemical potentials
at infinity. Secondly, we prove that the same formula can be obtained
through the anisotropic S-duality of the partition function of a dual
2D field theory, and we show its close relationship with the corresponding
Cardy-like formula. Finally, by considering the scaling properties
of the entropy as a function of the charges, it was possible to recover
the aforementioned Smarr relation from the Euler theorem for homogeneous
functions.

The second aim of this work is devoted to the thermodynamical stability
of the system. We have shown that, as it is expected for a black hole
solution in a Chern-Simons theory, the specific heat of the BTZ black
hole is a positive, monotonically increasing function of the temperature,
independently of the choice of KdV-type boundary condition. In contrast,
it was shown that the global stability of the system is sensitive
to the specific choice of boundary conditions. There is Hawking-Page
phase transition at an specific $z$-dependent self-dual temperature
$T_{s}$, for which, at temperatures below this point, the preferred
phase is the AdS$_{3}$ spacetime, and for higher temperatures, the
BTZ black hole is the more stable phase. This self-dual temperature
decreases for higher values of $z$, as does the size of the black
hole horizon and the energy that the system absorbs from the environment
in order for the transition occurs.

Remarkably, the anisotropic scaling properties which are commonly
realized in the context of Lifshitz holography, now take place in
the General Relativity scenario. This is because the KdV-type boundary
conditions induce these kind of scaling properties in the dual theory
allowing to study Lifshitz holography in a simple setup. This fact
leads to an interesting consequence, as it can be seen that rotation
terms naturally appears in the anisotropic Smarr formula, despite
that there is no a rotating Lifshitz black hole in three dimensions.

Along this work it has been assumed that the cosmological constant
is a fixed constant without variation. However, is it possible to
follow another point of view. If the energy of the black hole is no
longer the mass but the thermodynamical enthalpy, the Smarr formula
and a extended first law, can be found by considering a variable cosmological
constant which can be related with pressure and volume terms (see
e.g. \cite{PdV1,PdV2,PdV3,Mann}). In the literature, there is a standard
mechanism described in \cite{Henneaux:1984ji}, which explain how
to promote the cosmological constant to a canonical variable. Nonetheless,
at least in three dimensions, there is a superselection rule that
forbids this possibility \cite{Bunster:2014cna}, in consequence,
it cannot be rescaled.

\acknowledgments

We would like to thank Oscar Fuentealba, Cristián Martínez, Alfredo
Pérez, David Tempo and Jorge Zanelli, for helpful comments and discussions.
We are indebted to Ricardo Troncoso for especially valuable comments
and encouragement. C.E. thanks CONICYT through Becas Chile programme
for financial support. The work of P.R. was partially funded by PhD
CONICYT grant Nº 21161262 and Fondecyt grant Nº 1171162. M.R. is supported
by Fondecyt grant Nº 3170707. Centro de Estudios Científicos (CECs)
is funded by the Chilean Government through the Centers of Excellence
Base Financing Program of CONICYT.

\end{document}